\documentclass[a4paper]{article}
\usepackage{epsfig}
\begin{document}

\title{\bf\large Decay  $\phi\to\pi^+\pi^-$}

\author{
M.N.Achasov, 
K.I.Beloborodov,
A.V.Berdyugin,
A.V.Bozhenok, 
A.D.Bukin, \\
D.A.Bukin, 
\underline{S.V.Burdin\thanks{E-mail: burdin@inp.nsk.su}},
T.V.Dimova, 
V.P.Druzhinin, 
M.S.Dubrovin, \\
D.I.Ganyushin,
I.A.Gaponenko, 
V.B.Golubev,
V.N.Ivanchenko, \\ 
P.M.Ivanov, 
A.A.Korol, 
S.V.Koshuba,
E.V.Pakhtusova,\\
E.A.Perevedentsev, 
A.A.Salnikov, 
S.I.Serednyakov,
V.V.Shary, \\
Yu.M.Shatunov, 
V.A.Sidorov, 
Z.K.Silagadze, 
A.A.Valishev,\\
A.V.Vasiljev
\\ \\
{\it Budker Institute of Nuclear Physics and Novosibirsk State University} \\ 
{\it 630090, Novosibirsk, } \\ 
{\it Russia }
}

\date{}
\maketitle

\begin{abstract}
The process $e^+e^-\to\pi^+\pi^-$ 
has been studied with the SND detector at VEPP-2M
$e^+e^-$ collider in the vicinity of  $\phi(1020)$
resonance. 
From the analysis of the energy dependence of measured cross section
the branching ratio 
$B(\phi\to\pi^+\pi^-)=(7.1\pm 1.1\pm 0.9)\cdot 10^{-5}$ 
and the  phase $\psi_\pi=-(34\pm 4\pm 3)^\circ$ of interference 
with the non-resonant $\pi^+\pi^-$ production amplitude were obtained.
\end{abstract}

 \twocolumn

\section{Introduction}
 The decay $\phi\to \pi^+\pi^-$ reveals itself as an
interference pattern in the energy
dependence of the cross section of the process $e^+e^-\to\pi^+\pi^-$ 
in the region close to $\phi$ peak. 
The $\phi\to\pi^+\pi^-$ decay was previously
studied at VEPP-2M collider \cite{Olya,ND} and current
PDG value $B(\phi\to\pi^+\pi^-)=(8^{+5}_{-4})\cdot
10^{-5}$ \cite{PDG} is based on these results.

The decay $\phi\to\pi^+\pi^-$ violates both OZI rule and G-parity 
conservation. 
The decay
amplitude in Vector Dominance Model (VDM) was calculated in
\cite{Karn}. The main contribution into the amplitude of the 
$\phi\to\pi^+\pi^-$ decay in this work comes from the electromagnetic
$\phi-\rho$ mixing. The contribution of the $\phi-\rho$ transitions
through the $\omega$ meson and other intermediate states 
such as $K\bar{K}$, $\eta\gamma$, etc. is
estimated to be $\sim 20\%$ of electromagnetic
$\phi-\rho$ mixing.  The value of the branching ratio of the
decay $\phi\to\pi^+\pi^-$ calculated from the decay amplitude
obtained in the work \cite{Karn} is almost 2 times
higher than current PDG value \cite{PDG}.
Different $\phi-\omega$ mixing models were
scrutinized in respect to this decay in \cite{Ach1}. 
The branching ratio calculated in this work is lower than that in
 \cite{Karn}, but discrepancy between the experimental results,
 especially \cite{ND},  and the theoretical prediction \cite{Ach1}
still  exists.
Possible mechanisms, which could decrease the theoretical 
branching ratio, are discussed in \cite{Ach1}. One of them is
the existence of direct decay $\phi\to\pi^+\pi^-$. 

\section{Experiment}
\begin{figure}[htb]
\begin{minipage}[t]{0.475\textwidth}
\centerline{\includegraphics[width=8cm]{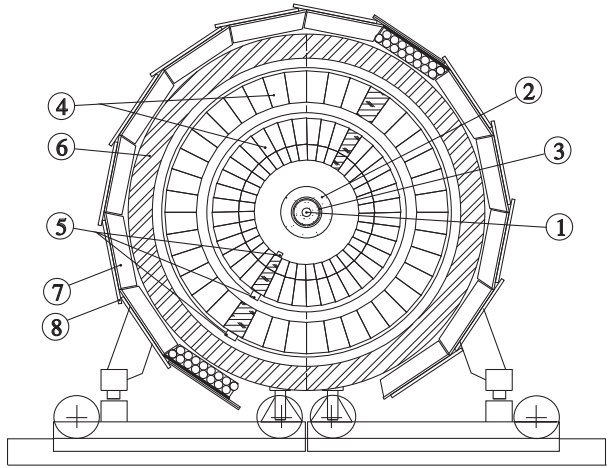}}
\caption{
Detector  SND  --- section 
across the beam;
1 --- beam pipe, 2 --- drift chambers, 3 --- inner scintillation
counters, 4 
--- NaI(Tl) counters, 5 --- vacuum phototriodes, 6 --- iron absorber, 7 
--- streamer tubes, 8 --- outer scintillation counters
}
\label{SNDtrans}
\end{minipage}
\end{figure}

The experiments with SND detector (Fig.~\ref{SNDtrans}) at 
VEPP-2M $e^+e^-$ collider are being conducted since 1995. 
SND is a general purpose non-magnetic
detector \cite{SND}. The main part of the SND is a 3-layer spherical
electromagnetic calorimeter, consisting of 1632 NaI(Tl)
crystals \cite{Lisbon}. 
The solid angle of the calorimeter is $\sim 90\%$ of $4\pi$
steradian. The angles of charged particles  are
measured by two cylindrical drift chambers covering 95\% of full
solid angle. The important part of the detector for the
process under study is the outer muon system, consisting of streamer tubes
and plastic scintillation counters.

  The 1998 experiment was carried out in the energy range
$2\mathrm{E}_b$=984--1060~MeV in 16 energy points and consisted of 2
data taking runs  \cite{Prep.98}: PHI\_9801, PHI\_9802. 
The total integrated luminosity $\Delta L=8.6$~pb$^{-1}$ collected in
these runs corresponds to  $13.2\cdot 10^{6}$
produced $\phi$ mesons.
The integrated
luminosity was measured using $e^+e^-\to e^+e^-$ events selected in
the same acceptance angle as the events of the process under study. 
The interference term in the $e^+e^-\to e^+e^-$ cross section
due to $\phi\to e^+e^-$ decay was also taken into account.
The systematic error of the luminosity measurement was estimated to be 
 2\%. 

\section{Event selection} 
  The energy dependence of the cross section of the process 
\begin{equation}
 e^{+}e^{-} \to\pi^+\pi^-
\label{pipi}
\end{equation}
was studied in the vicinity of $\phi$ meson. Events 
containing two collinear charged particles and no photons were
selected for analysis.
The following cuts on  angles of acollinearity
 of the charged particles in azimuthal and polar directions were imposed:
$\mid\Delta\varphi\mid<10^\circ$, $\mid\Delta\theta\mid<25^\circ$. 
To suppress the beam background the production point of charged
particles was required to be within $0.5$~cm from the 
interaction point in the azimuthal plane and $\pm 7.5$~cm along 
the beam direction (the longitudinal size of the interaction region 
 $\sigma_{z}$ is about $2$~cm). 
The polar angles of the charged particles were 
required to be in the range 
$45^\circ<\theta<135^\circ$, determined by acceptance angle of 
the muon system.

\begin{figure}[htb]
\begin{minipage}[htb]{0.49\textwidth}
\centerline{\includegraphics[width=8cm]{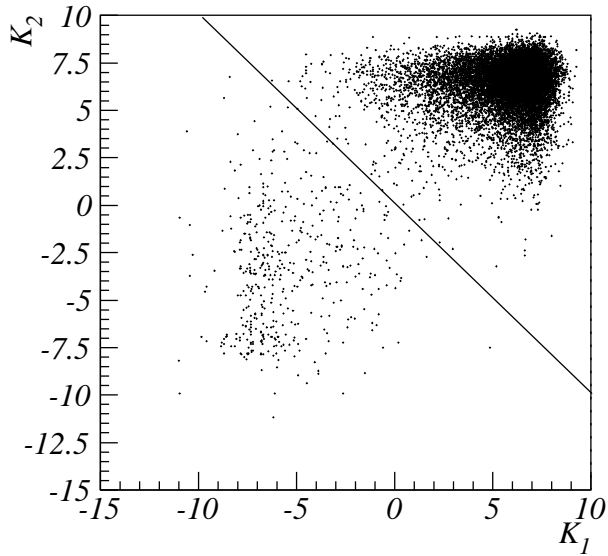}}
\caption{ 
Distribution of the parameters  $K_1$ and $K_2$ for two electrons
and  pions from the processes (\ref{ee}) and (\ref{pipi}). 
Electrons occupy the top right corner while pions concentrate in
the bottom left one.
}
\label{epi}
\end{minipage}
\end{figure}

The main sources of background are cosmic muons and the following
 processes:
\begin{eqnarray}
 e^{+}e^{-} \to e^{+}e^{-},
\label{ee} \\
 e^{+}e^{-} \to\mu^+\mu^-, 
\label{mumu} \\
 e^{+}e^{-} \to \pi^{+}\pi^{-}\pi^0, 
\label{ppp} \\
 e^{+}e^{-} \to K_{S}K_{L}.
\label{KsKl} 
\end{eqnarray}
To suppress the background from the process (\ref{ee}) a procedure of
$e/\pi$ separation was used. It utilizes the difference in the 
longitudinal energy deposition profiles in the calorimeter
for electrons and pions. The separation parameter was calculated
for each charged particle
in an event: 
\begin{equation}
K=\log\left(\frac{\mathcal{P}_e(E_1,E_2,E_3,E_e)}
{\mathcal{P}_\pi(E_1,E_2,E_3,E_\pi)}\right),
\label{epif} 
\end{equation}
where $\mathcal{P}_{e(\pi)}$ --- the probability for
an electron (pion) with the energy $E_{e(\pi)}$ 
to deposit the energy  $E_i$ in the $i$-th calorimeter layer.
$E_{e(\pi)}$ in our case is equal to the beam energy.
The separation parameters distribution
for both particles in collinear events with no hits in
the muon system, is shown in Fig.~\ref{epi}. This distribution is
asymmetric because the particles are ordered
according to their energy depositions in the calorimeter. 
To select the events of the process (\ref{pipi}) the cut $K_1+K_2<0$
was imposed. The background from the process (\ref{ee}) was suppressed by
a factor of $\sim 3000$, while only 7\% of the events of the
process under study were lost. Remaining background from the
process $e^+e^-\to e^+e^-$ was about 1.5\%.

The events of the process (\ref{mumu}) and cosmic muons can be
efficiently  suppressed by the muon system. We required
no hits in the scintillation counters of the muon
system. The efficiency of these counters was estimated
using cosmic muons selected by special cuts.
Due to possible admixture of beam events which actually produce no
hits in the muon counters only the lower boundary
of the efficiency was obtained: 99.8\%. Thus
estimated contribution of cosmic muons does not exceed 0.7\% 
of the total number of events of the process (\ref{pipi}) and was neglected.

The energy dependences of the
probabilities for muons and pions to produce hits in outer scintillation
counters were obtained from the experimental data.
With energy increasing from 492~MeV up to 530~MeV these probabilities
rise from  84\% up to 94\%  for muons and from 0.5\% to 11\% for pions.  
In the final selection of the process (\ref{pipi}) the background
from the process (\ref{mumu}) was about 15\%. 

To suppress the resonant background from the processes (\ref{ppp})
 and (\ref{KsKl}) the following cuts on energy depositions in the
 calorimeter were applied: 
\begin{enumerate}
\item the energy deposition in the first calorimeter layer 
 of the most energetic particle in an event is less than 75~MeV;
\item the energy deposition in the third calorimeter layer
 of the least energetic particle in an event is more than 50~MeV.
\end{enumerate}
In the events of the process (\ref{ppp}), which satisfy the
 geometrical cuts,  the energetic photon from $\pi^0$ decay 
propagates along the direction of a charged pion producing
unusually large energy deposition in
 the first calorimeter layer for this pion. Such
 events are suppressed by the first cut. The
 pions from process  (\ref{KsKl}) are relatively soft with a maximum
 energy of about 300~MeV and low probability of significant energy
 deposition in the third calorimeter layer. The second cut is crucial
 for the rejection of the process  (\ref{KsKl}).
 The residual cross sections of resonant background processes were estimated
 by Monte Carlo (MC) simulation: 0.06~nb for 
 the process (\ref{ppp}), 0.09~nb for the process (\ref{KsKl}).

 To determine the remaining resonant background more accurately 
 the selected events were divided into two data samples
 using the parameter $\Delta\varphi$: 
 $\mid\Delta\varphi\mid<5^\circ$ and
 $\mid\Delta\varphi\mid>5^\circ$. The resolution in $\Delta\varphi$ is 
 about $1^\circ$. The main part of the events of the process
 (\ref{pipi}) is contained in the first sample. Due to the emission of
 hard photons by initial or final particles and errors in
 the  reconstruction of the particle angles
 some events of the process (\ref{pipi}) can migrate into
 the second sample. The level of the resonant cross section
 $\sigma^{res}_2$, determined in the second sample, was used to
 estimate the resonant background in the first sample:
 $\sigma_1^{res}=k\sigma_2^{res}$.
 The coefficient $k=1.5\pm0.3$
 was obtained by MC simulation of the processes (\ref{ppp})
 and (\ref{KsKl}), its error  is determined by accuracy of simulation
 of energy depositions of pions in the calorimeter.
 Because the level of the resonant
 background is low, the error in $k$ does not give significant
 contribution into the errors of the interference parameters. 
 The cut
 $\mid\Delta\varphi\mid<5^\circ$ reduces the level of resonant background down
 to as low as 0.09~nb. This value is less than 1\% of the process
 (\ref{pipi})  detection cross section.

\begin{figure}[htb]
\begin{minipage}[htb]{0.49\textwidth}
\centerline{\includegraphics[width=8cm]{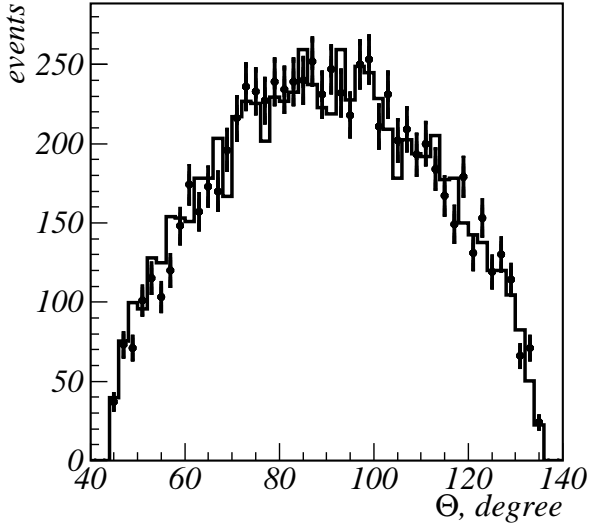}}
\caption{ 
The pion polar angle distributions for the experimental events (points
with errors) and simulation (histogram) for the process
$e^+e^-\to\pi^+\pi^-$.  
}
\label{teta2}
\end{minipage}
\end{figure}

 The pion polar angle  distribution
 for the  process (\ref{pipi}) at beam energy
 higher than 520~MeV is shown in Fig. \ref{teta2}. 
 At this energy the cross sections of the resonant processes (\ref{ppp})
 and (\ref{KsKl}) are small. The additional cut on the total energy
 deposition in the calorimeter $E_{tot}>400$~MeV rejects the events of
 the process (\ref{mumu}). A good agreement between experimental
 distribution and the simulation of the process (\ref{pipi}) shows that
 selected pion sample is quite pure and the level of QED background is low.

\section{Data analysis}
\begin{figure}[htb]
\begin{minipage}[t]{0.45\textwidth}
\centerline{\includegraphics[width=8cm]{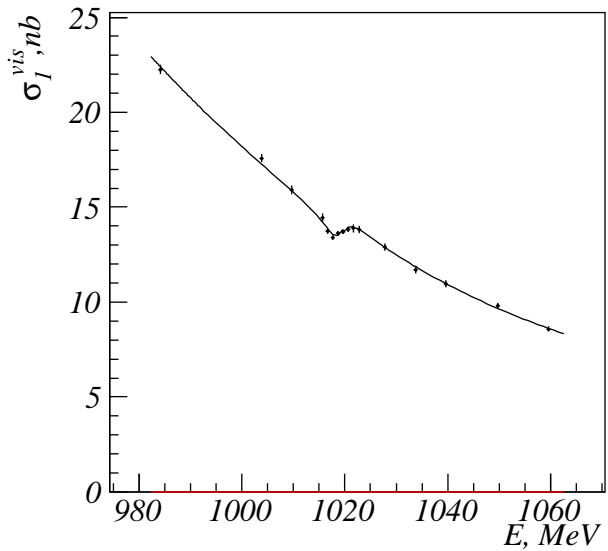}}
\caption{  The detection cross section in the first data sample
 \mbox{($\mid \Delta\varphi\mid <5^\circ$)}.
}
\label{pipitot}
\end{minipage}
\end{figure}

\begin{figure}[htb]
\begin{minipage}[t]{0.45\textwidth}
\centerline{\includegraphics[width=8cm]{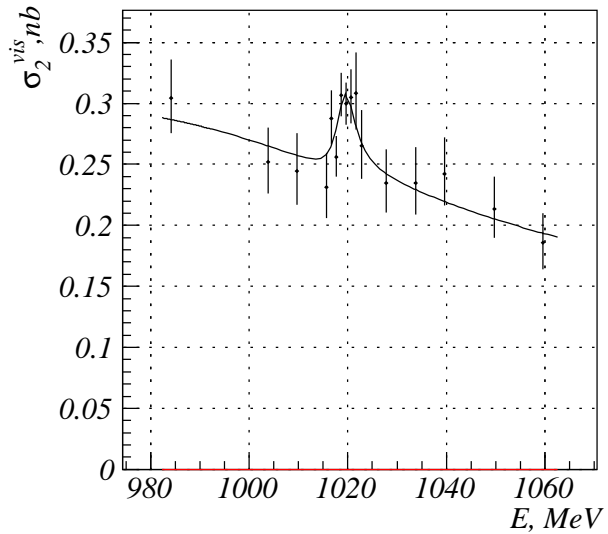}}
\caption{  The detection cross section in the second data sample 
 \mbox{($\mid\Delta\varphi\mid>5^\circ$)}. 
}
\label{restot}
\end{minipage}
\end{figure}

The fitting of the detection cross sections for the first 
  and the second 
  samples were 
performed simultaneously (Fig.~\ref{pipitot},~\ref{restot}).  To describe
the cross sections the following formulae were used:  
\begin{eqnarray}
\label{cross1}
\sigma^{vis}_1(E) & = & \sigma^{vis}_{\pi\pi}(E)  
+\sigma^{vis}_{\mu\mu}(E) \nonumber \\
& & +\sigma^{vis}_{ee}(E)+\sigma^{res}_1(E), \\
\label{cross2}
\sigma^{vis}_2(E) & = & C+D\cdot(E-m_\phi) \nonumber \\
& & +\sigma^{res}_2(E), \\ 
\sigma^{res}_1(E) & = & k\sigma^{res}_2(E), \nonumber \\
\label{rezfon}
\sigma^{res}_2(E) & = & \varepsilon_{res} \cdot 
(0.39\cdot\sigma_{\pi^+\pi^-\pi^0}(E) \nonumber \\
& & +  0.61\cdot\sigma_{K_S K_L}(E)), \nonumber
\end{eqnarray}
where $E$ is the CM energy; $\sigma^{vis}_{\pi\pi}(E)$ ---
the detection cross section of the process (\ref{pipi});
$\sigma^{vis}_{\mu\mu}(E)$ --- the contribution of the process
(\ref{mumu}) (this process was studied in our work \cite{mymumu}); 
$\sigma^{vis}_{ee}(E)=0.2(nb)\cdot (m_\phi/E)^2$ --- the contribution
of the process (\ref{ee}).  
The ratio 0.39:0.61 between the processes (\ref{ppp}) and
(\ref{KsKl}) was taken from the simulation.
The coefficients $C$, $D$ and $\varepsilon_{res}$ were free fit parameters.

The following expression was used for $\sigma^{vis}_{\pi\pi}$: 
\begin{eqnarray}
 \sigma^{vis}_{\pi\pi}(E) & = &\sigma_0(E)\cdot R(E)\left| 1 -
 Z_\pi\frac{m_{\phi}\Gamma_{\phi}}{\Delta_{\phi}(E)}\right|^2, \nonumber \\
\label{crosspi} 
 \sigma_0(E) & = & \frac{\pi\alpha^2\beta^3(E)\mid F_\pi(E) \mid^2}{3\cdot
E^2}, 
\end{eqnarray}
where $\alpha$ is the fine structure constant;
$\beta(E)=(1-4\cdot m^2_\pi/E^2)^{1/2}$;
$m_\phi$, $\Gamma_{\phi}$,
$\Delta_{\phi}(E)
= m_{\phi}^{2} - E^2 - iE\Gamma(E)$ ---
$\phi$-meson mass, width and  propagator respectively; 
$\sigma_0(E)$ --- the Born
cross section of the process $e^+e^-\to\pi^+\pi^-$;
$Z_\pi$ --- complex parameter characterizing strength of the
interference.
Two representations of $Z_\pi$ are used in different works: 
$Z_\pi=Q_\pi e^{i\psi_\pi}=\mathbf{Re} Z_\pi+i\mathbf{Im} Z_\pi$. 
$F_\pi(E)$ is the pion
form factor without $\phi$-meson contribution:
\begin{equation}
\label{piform}
\mid F_\pi(E) \mid^2=\mid F^\phi_\pi\mid^2\cdot
(1+A\cdot(E-m_\phi)+B\cdot(E-m_\phi)^2), 
\end{equation}
with $F^\phi_\pi$ as the pion form factor at the maximum of $\phi$
resonance. $Q_\pi$, $\psi_\pi$, $A$, $B$ and $\mid F^\phi_\pi\mid^2$ 
are free fitting parameters.
 $R(E)$ is a factor taking into account
detection efficiency and radiative corrections:
\begin{equation}
R(E)=\varepsilon_\pi\frac{\sigma_{\pi\pi}(E)}{\sigma'_0(E)
\left|  1 - Q'_\pi
e^{i\psi'_\pi}\frac{m_{\phi}\Gamma_{\phi}}{\Delta_{\phi}(E)} \right| ^2}.
\label{Rfac}
\end{equation}
 $\sigma_{\pi\pi}$ is the result of MC integration of differential
cross section of the process (\ref{pipi}) with all geometrical restrictions
\cite{theor}.
Since the probability for pions to hit the outer scintillation counters
depends on energy, it was taken into account during
$\sigma_{\pi\pi}$ calculation. The remaining contributions into the
detection efficiency do not depend on CM energy and pions energies
and were included into $\varepsilon_\pi$.
The value $\varepsilon_\pi=0.234$ was obtained using MC simulation and 
experimental data. It is mainly determined by the cuts on energy depositions. 
Its independence of the pions energy was checked
in the range 430 -- 530~MeV  using the
pions from the process (\ref{ppp}) with energies up to 450~MeV and
pions from the process (\ref{pipi}) at the beam energy 530~MeV.
The geometrical cuts and the requirement on no hits
in the outer scintillation counters led to 
50\% efficiency losses,
so the total detection efficiency of the process (\ref{pipi}) was
approximately 12\% at $E=m_\phi$.

  The $R(E)$ was calculated by iteration method. As a first
approximation the interference parameters $Q'_\pi$ and $\psi'_\pi$ from \cite{ND} 
were used.
The pion form factor was taken from 
 \cite{formpi}  while calculating $\sigma'_0(E)$. 
After fitting $R(E)$ was recalculated
  with corrected $Q'_\pi$ and $\psi'_\pi$. This procedure was repeated
  until convergence was reached.

The branching ratio $B(\phi\to\pi^+\pi^-)$ is related to the
interference parameters by the following formula:
\begin{equation}
 B(\phi\to\pi^+\pi^-)=\frac{Q_\pi^2\alpha^2\beta^3(m_\phi) \mid
                           F^\phi_\pi\mid^2} 
                           {36\cdot B(\phi\to e^+ e^-)}, 
\label{Bpipi} 
\end{equation}
where  
 $B(\phi\to e^+ e^-)=(2.99\pm 0.08)\cdot 10^{-4}$ \cite{PDG}.

 The fitting has been performed for each experimental run separately. The
 results are listed in Table \ref{tabres}. The fit parameters
 for two runs are in statistical agreement, 
 therefore combined fit was performed to
 obtain the final results also listed in Table \ref{tabres}.
The observed
 level of resonant background 0.07~nb is in a good agreement with
 the MC estimation of 0.09~nb.
 The fitted values of the coefficients $A$ and $B$ from the equation 
 (\ref{piform}) are $A=-(8.5\pm0.3)\cdot 10^{-3}\
 MeV^{-1}$ and $B=(4.9\pm 1.0)\cdot 10^{-5}\ MeV^{-2}$.

To check the accuracy of the process (\ref{mumu}) background
 subtraction, the fit to the data with more stringent event selection 
cuts has been
done. The additional requirement that the total energy deposition in the
calorimeter is higher than 400~MeV significantly reduced the muon
background. The obtained  interference parameters: 
\begin{eqnarray}
 Q_\pi = 0.073\pm0.006, \nonumber \\
 \psi_\pi = -(32\pm 5)^\circ \nonumber  
\end{eqnarray}
agrees well with the results from the Table~\ref{tabres}.

The additional contribution into the shape of interference pattern
may come from the process 
\begin{equation}
e^+e^-\to\phi\to f_0\gamma\to \pi^+\pi^-\gamma. 
\label{f0g}
\end{equation}
The process (\ref{f0g}) interferes with the process (\ref{pipi}) when
soft photon is emitted by pions.
This contribution estimated using 
CMD-2 analysis of the process $e^+e^-\to\pi^+\pi^-\gamma$ in the vicinity
of $\phi$ resonance
\cite{CMDppg} does not exceed 1.5\% of the
interference under study. This value was included into the systematic
error. 
\begin{table*}[htb]
\large
\begin{center}
\caption
{
The fit results for two runs and the
combined fit.
}
\label{tabres}
\begin{tabular}{|c|c|c|c|}
\hline
Parameter & PHI\_9801   &  PHI\_9802  & Combined \\
\hline
$\chi^2/NDF$ & $15.3/26$ &  $ 27.5/26$   &  $48.4/58$      \\
$\mid F^\phi_\pi \mid^2$ & $2.96\pm 0.03$ & $3.01\pm 0.03$ & $2.98\pm 0.02$ \\
$Q_\pi$ & $0.069 \pm 0.008$ & $0.078\pm 0.007$ & $0.073\pm 0.005$\\
$\psi_\pi$ & $ -(33\pm 7)^\circ$ & $ -(35\pm 6)^\circ$ & $ -(34\pm 4)^\circ$\\
$B(\phi\to\pi^+\pi^-)$& $ (6.2\pm1.4)\cdot 10^{-5}$ 
& $ (8.2\pm1.8)\cdot 10^{-5}$ & $ (7.1\pm1.1)\cdot 10^{-5}$ \\
\hline
\end{tabular}
\end{center}
\end{table*}

 The representation
$Z_\pi=\mathbf{Re} Z_\pi+i\mathbf{Im} Z_\pi$ is
suitable to present the different contributions into the systematic error
of the interference parameters:
\begin{enumerate}
\item the calculation of the radiative corrections:
$\mathbf{Re} Z_\pi$ ---  5\%, $\mathbf{Im} Z_\pi$ ---  3\%;
\item the subtraction of the non-resonant background:
$\mathbf{Re} Z_\pi$ ---  0.8\%, $\mathbf{Im} Z_\pi$ ---  0.6\%;
\item the contribution of the process (\ref{f0g}):
$\mathbf{Re} Z_\pi$ ---  1.5\%, $\mathbf{Im} Z_\pi$ ---  1.5\%;
\item the model dependence on the choice of the function
approximating the pion form factor:
$\mathbf{Re} Z_\pi$ ---  1\%, $\mathbf{Im} Z_\pi$ ---  8\%;
\item the subtraction of the resonant background:
$\mathbf{Im} Z_\pi$ --- 3\%.
\end{enumerate}

The systematic error of $\mid F^\phi_\pi\mid^2$ is
determined by the error of the detection efficiency $\varepsilon_\pi$
\mbox{($\sim$ 
5\%)} and the accuracy of luminosity determination (2\%).

 The final results are the following:
\begin{eqnarray}
\label{result}
 \mid F^\phi_\pi \mid^2 & = & 2.98\pm0.02\pm0.16,  \\
 Q_\pi & = & 0.073\pm0.005\pm0.004, \nonumber \\
 \psi_\pi & = & -(34\pm4\pm3)^\circ, \nonumber \\
 B(\phi\to\pi^+\pi^-) & = & (7.1\pm 1.1 \pm 0.9)\cdot
10^{-5}. \nonumber 
\end{eqnarray}
 For another representation of $Z_\pi$ we obtained: 
\begin{eqnarray}
\mathbf{Re} Z_\pi & = &\quad  0.061\pm 0.005 \pm 0.003, \nonumber  \\
\mathbf{Im} Z_\pi & = &- 0.041\pm 0.006 \pm 0.004. \nonumber 
\end{eqnarray}

\section{Discussion}

 The obtained value of the branching ratio
 $$B(\phi\to\pi^+\pi^-)=(7.1\pm 1.1\pm 0.9)\cdot 10^{-5}$$
 agrees well
 with the world average value $B(\phi\to\pi^+\pi^-)=(8^{+5}_{-4})\cdot
 10^{-5}$ \cite{PDG} and has a 3 times higher accuracy. However there is
 a discrepancy between our result and the preliminary
 result of CMD-2 experiment \cite{CMD2PI}:
$B(\phi\to\pi^+\pi^-) =  (18.1\pm 2.5 \pm 1.9)\cdot 10^{-5}.$

 The measured value $\mathbf{Im} Z_\pi=- 0.041\pm 0.006 \pm 0.004$ 
agrees with the theoretical predictions \cite{Karn} while 
the value $\mathbf{Re} Z_\pi = 0.061\pm 0.005 \pm 0.003$ 
is  2.5 times lower than the expected value. 
The different models of the $\phi-\omega$
mixing were examined in the work \cite{Ach1}. The lowest value 
$\mathbf{Re} Z^{th}_\pi = 0.12$ from this work also contradicts our
results.
This disagreement could 
be understood if the direct decay $\phi\to\pi^+\pi^-$ exists or/and in case
of nonstandard $\rho-\omega-\phi$ mixing. One can notice that the
measured branching ratio of another rare decay $\phi\to\omega\pi^0$
\cite{omegapi}, which violates OZI rule and G-parity,
disagrees with theoretical predictions.

\section{Acknowledgement}
This work is supported in part by Russian Fund for basic 
researches (grant 99-02-16815) and 
STP "Integration" (No.274).

\end{document}